\documentclass[apjl]{emulateapj}
\usepackage{graphicx,subfigure,amsmath, amsfonts, amssymb,aas_macros,footnote,color,times,natbib}
\newcommand{\kms}{km~s$^{-1}$}

\newcommand{\twco}{\hbox{$^{12}$CO}}

\newcommand{\hi}{\hbox{H\,{\sc i}}}

\shorttitle{Resolved Molecular Outflow in Mrk\,231}
\shortauthors{K. Alatalo}
\slugcomment{Accepted by ApJ Letters}

\begin{document}


\title{Escape, Accretion or Star Formation?  The Competing Depleters of Gas in the Quasar Markarian 231}

\author{Katherine Alatalo$^{1}$}
\affil{$^1$Infrared Processing \& Analysis Center,  California Institute of Technology, Pasadena, CA 91125, USA
}


\begin{abstract} 
We report on high resolution CO(1--0), CS(2--1) and 3mm continuum Combined Array for Research in Millimeter Astronomy (CARMA) observations of the molecular outflow host and nearest quasar Markarian 231.  We use the CS(2--1) measurements to derive a dense gas mass within Mrk\,231 of $1.8\pm0.3\times10^{10}~M_\odot$, consistent with previous measurements.  The CS(2--1) data also seem to indicate that the molecular disk of Mrk\,231 is forming stars at about normal efficiency.  The high resolution CARMA observations were able to resolve the CO(1--0) outflow into two distinct lobes, allowing for a size estimate to be made and further constraining the molecular outflow dynamical time, further constraining the molecular gas escape rate.  We find that 15\% of the molecular gas within the Mrk\,231 outflow actually exceeds the escape velocity in the central kiloparsec. Assuming that molecular gas is not constantly being accelerated, we find the depletion timescale of molecular gas in Mrk\,231 to be 49\,Myr, rather than 32\,Myr, more consistent with the poststarburst stellar population observed in the system.
\end{abstract}

\keywords{galaxies: active -- galaxies: evolution -- galaxies: ISM -- galaxies: nuclei -- galaxies: quasars: individual (Markarian 231)}

\section{Introduction}
Understanding the role of active galactic nuclei (AGN) in its host galaxy's transition from a blue starforming disk to a red quiescent ellipsoid is one of the forefront problems in studying galaxy evolution, with molecular gas at the center of the debate.  A dearth of galaxies appearing in the optical ``green valley'' \citep{faber+07} seems to indicate that galaxies must transition between these two stages rapidly.  It has been suggested that the expulsion of a galaxy's starforming material through AGN feedback might provide a mechanism that is both sufficiently powerful and rapid to achieve this transition \citep{springel+05,hopkins+06}.  Recent observations have shown many galaxies with AGNs host molecular outflows (\citealt{feruglio+10,alatalo+11,aalto_1377,cicone+14} and references therein).  In fact, AGN-driven molecular winds might be a ubiquitous property, as molecular observations of these types of objects become more numerous.


Markarian 231 (at a distance of 181\,Mpc; \citealt{boksenberg+77}) is an ideal laboratory to study the ways in which supermassive black holes and star formation interact with molecular gas.  Mrk\,231 has both a rapidly accreting, powerful AGN \citep{braito+04} and a starburst, with a star formation rate of $170~M_\odot$~yr$^{-1}$ \citep{veilleux+09}, and is classified an ultraluminous infrared galaxy (ULIRG; \citealt{sanders+mirabel96}).  Its disturbed morphology \citep{hutchings+neff87} confirm that it is in the late stages of an ongoing merger, which \citet{hamilton+keel87} suggest is the origin of both the starburst and quasar activity.  Recently, it has also been shown that Mrk\,231 hosts a massive outflow, observed in molecular gas \citep{fischer+10,feruglio+10,aalto+12}, neutral gas \citep{rupke+05,rupke+11,teng+13} and warm ionized gas \citep{lipari+09,rupke+11,veilleux+13}.  These properties make Mrk\,231 the ideal laboratory to test the ways in which dense molecular gas, star formation, and an AGN intermingle within a merging system, shedding light on the fate of the starforming material, and ultimately, the galaxy as a whole, after a merger.

We present new high resolution Combined Array for Research in Millimeter Astronomy (CARMA; \citealt{carma}) continuum, CO(1--0) and CS(2--1) observations of Mrk\,231.  In \S\ref{obs}, we describe the observations and data analysis.  In \S\ref{disc}, we use the CARMA measurements to put the molecular gas in Mrk\,231 into context, both in terms of star formation properties and depletion timescales.  In \S\ref{conc}, we present our conclusions.  We use the cosmological parameters $H_0 = 70$~km~s$^{-1}$, $\Omega_M = 0.3$ and $\Lambda = 0.7$ \citep{wmap} throughout.

\section{Observations and Data Reduction}
\label{obs}
\begin{figure}[t]\centering
\includegraphics[width=0.49\textwidth]{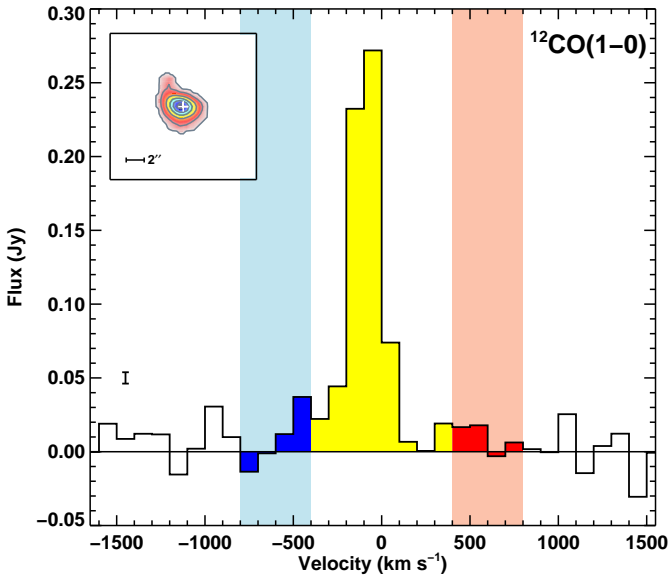}
\caption{Integrated CO(1--0) spectrum of Mrk\,231 from CARMA.  The shaded regions correspond to the ranges summed to search for wing emission, of $400<v<800$\,km~s$^{-1}$ for the red-shifted wing and $-800<v<-400$\,km~s$^{-1}$ for the blue-shifted wing.  The moment0 map is shown as an inset in the top left, where a white cross pinpoints the location of the AGN. }
\label{fig:cospec}
\end{figure}

Mrk\,231 was observed with CARMA between 2010 November and 2011 May in two different configurations: C array (1$''$ resolution) and B array (0\farcs7).  \twco(1--0) and CS(2--1) were observed simultaneously, utilizing the upgraded correlator.  The primary beam has a diameter of 2$'$ at CO(1--0), which covers all emission from Mrk\,231.  All observations used a long integration on a bright quasar to calibrate the passband, and alternated integrations between a gain calibrator \hbox{(1419+543)} and Mrk\,231.  We then used the Mrk\,231 3mm continuum source \citep{joyce+75} to self calibrate.  The self calibration step was able to mitigate atmospheric fluctuations for the longer baseline arrays.  The data were reduced using the Multichannel Image Reconstruction Image Analysis and Display ({\sc miriad}) software package \citep{miriad}. Calibration and data reduction steps were followed identically to \citet{alatalo+13}.   Figure \ref{fig:cospec} presents the CO(1--0) spectrum, using the aperture defined by the integrated intensity (moment0) map (seen as an inset in Fig. \ref{fig:cospec}). Table \ref{tab:results} presents the derived properties from the CARMA observations.

\begin{table*}
\centering
\caption{Mrk\,231 properties} \vspace{1mm}
\begin{tabular}{l c c c c c c c c}
\hline \hline
{\bf Tracer} & $\theta_{\rm maj}\times\theta_{\rm min}$ & K per Jy$^\ddagger$ & $\Delta v$ & RMS & RMS per channel & Area & $F_{\rm peak}$ & $F_{\rm int}$ \\
& ($''$) & & (km~s$^{-1}$) & (mJy~beam$^{-1}$) & (mJy) & $\Box''$ & (Jy~beam$^{-1}$~km~s$^{-1}$) & (Jy~km~s$^{-1}$) \\
\hline
{\bf CO(1--0)} & $1.80\times1.57$ & 35.54 & 100 & 2.77 & 7.70 & 7.7 & $30.4\pm5.8$ & $74.3\pm3.0$ \\
{\bf CS(2--1)} & $1.40\times1.26$  & 78.35 & 34 & 1.28 & 1.82 & 3.9 & $0.92\pm0.13$ & $1.43\pm0.19$ \\
\hline
{\bf Blue} & $1.19\times1.07$  & 78.48 & 400 & 0.57 & -- & 5.3 & $1.91\pm0.23$ & $4.30\pm0.53$ \\
{\bf Red} & $1.19\times1.07$  & 78.48 & 400 & 0.54 & -- & 5.9 & $1.50\pm0.21$ & $4.60\pm0.53$ \\
\hline \hline
\end{tabular} \\
\noindent $^\ddagger$ Kelvin per Jansky factor for each observation
\label{tab:results}
\end{table*}

The {\sc miriad} task {\tt uvlin} was used to separate out continuum emission from Mrk\,231 from the line emission, and we estimate the continuum flux of Mrk\,231 to be $20.9\pm0.3$ mJy\footnote{This does not include the absolute flux calibration uncertainty of 20\%} in the unself-calibrated data.  A centroid was computed to determine the position of the 3mm continuum and compare it to the VLBI-determined radio core \citep{ma+98}, and found that the two positions agree to 0.04$''$.

\begin{figure*}[t]\centering
\subfigure{\includegraphics[height=3.7in]{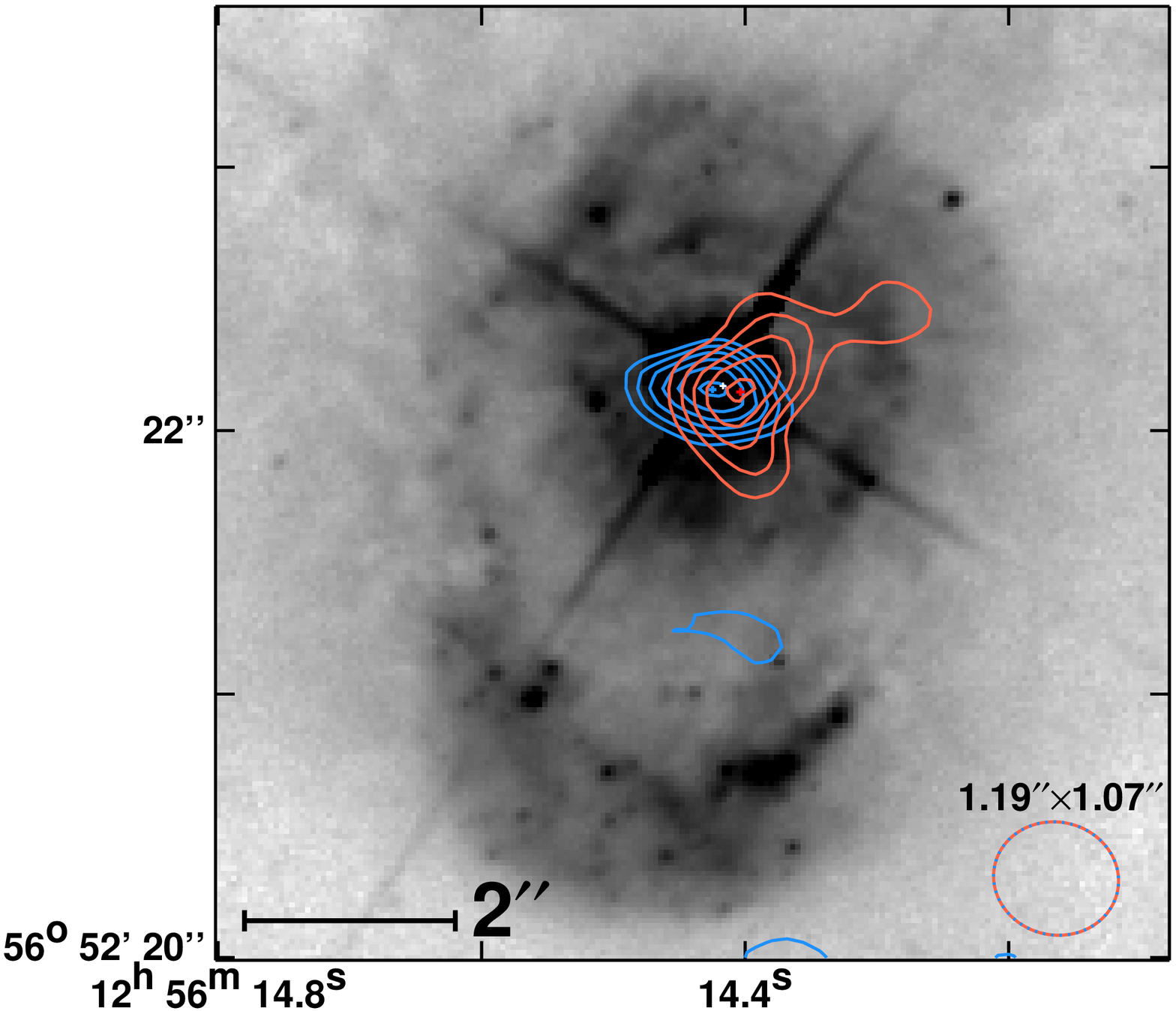}}
\subfigure{\includegraphics[height=3.7in,clip,trim=0cm 0.9cm 0cm 1cm]{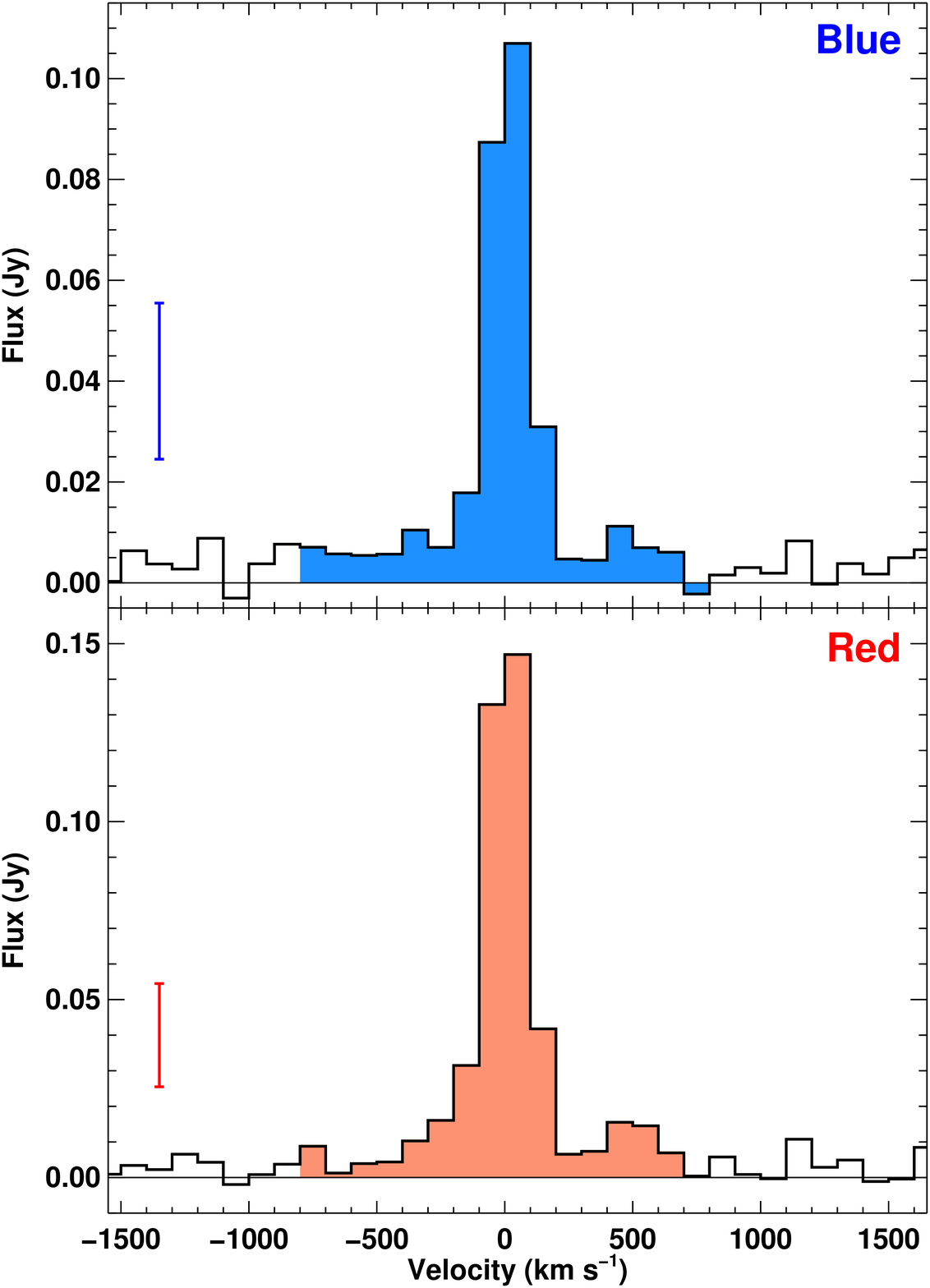}}
\caption{{\bf(Left):}  {\em B}-band archival image of Mrk\,231 from {\em HST} (grayscale, \citealt{kim+13}) is overlaid with the integrated CARMA maps of the blue-shifted ($-800<v<-400$ km~s$^{-1}$; blue) and red-shifted ($400<v<800$ km~s$^{-1}$; red).  Contours for the red lobe is [3,4,5,6,7]$\sigma$ and [3,4,5,6,7,8]$\sigma$ for the blue lobe, where $\sigma$ is the root mean square noise of the map.  The blue and red crosses represent the centroids on the blue- and red-shifted wings, respectively.  The white cross represents the position of the centroid of the 3mm continuum point source.  Separation between the lobes is $\approx 0.49''$, a well-resolved distance, where the size of the red and blue crosses represent the centroiding errors for each wing.  The separation is detected to 5.8 times the centroiding error.  It is of note that the CO(1--0) wings velocity range almost entirely exceeds the escape velocity out of the central kiloparsec of Mrk\,231. {\bf(Right):} Integrated CO(1--0) spectrum of Mrk\,231 using emission exceeding a signal-to-noise of 3 in red- and blue-shifted wing images to create the extraction apertures.  This seems to indicate that CARMA has successfully detected emission in the red- and blue-shifted wings.}
\label{fig:lobes}
\end{figure*}

To image the blue-shifted and red-shifted wings, we created images with velocities between $-800<v<-400$\,\kms\ and $400<v<800$\,\kms, respectively.  Figure \ref{fig:cospec} shows the continuum-subtracted CO(1--0) spectrum derived from the 100\,km~s$^{-1}$ CARMA channel maps, using the moment0 map (Fig. \ref{fig:cospec} inset) as the masking aperture in each channel. The corresponding wing images and spectra are shown in Figure \ref{fig:lobes}, overlaid on {\em B}-band {\em HST} imaging of Mrk\,231 (right; \citealt{kim+13}).  CARMA derives a total flux of $74.3\pm3.0$~Jy~\kms, in agreement within errors with the flux recovered by Plateau de Bure \citep{feruglio+10,cicone+12}.

CARMA has successfully resolved the high velocity CO(1--0) wings, finding a separation between the centroids of $0.49''$ (415\,pc), a distance 5.8 times larger than the centroiding error\footnote{$\epsilon_{\rm centroid} = \theta_{\rm beam}/(2\times{\rm SNR})$}, and the blue- and red-shifted lobes have sizes of 2.3$''$ (1980\,pc) and 2.2$''$ (1820\,pc), respectively.  The fluxes associated with each imaged wing was determined by summing the emission that lay inside the 3$\sigma$ contours of the emission, and were found to be $F_{\rm CO,blue} = 4.30\pm 0.53$~Jy~\kms and $F_{\rm CO,red} = 4.60\pm0.53$~Jy~\kms, which is larger than what was found in \citet{cicone+12}, though it is possible that our apertures were larger.  The lobes also appear to be separated in the East-West direction, with the centroid of the blue lobe located at [12:56:14.245, +56:52:25.24] and for the red lobe centroid at [12:56:14.213, +56:52:25.21], with the 3mm continuum point source (white cross) located between the lobes.

\begin{figure*}\centering
\subfigure{\includegraphics[height=4.8cm,clip,trim=0cm 0cm 0cm 0cm]{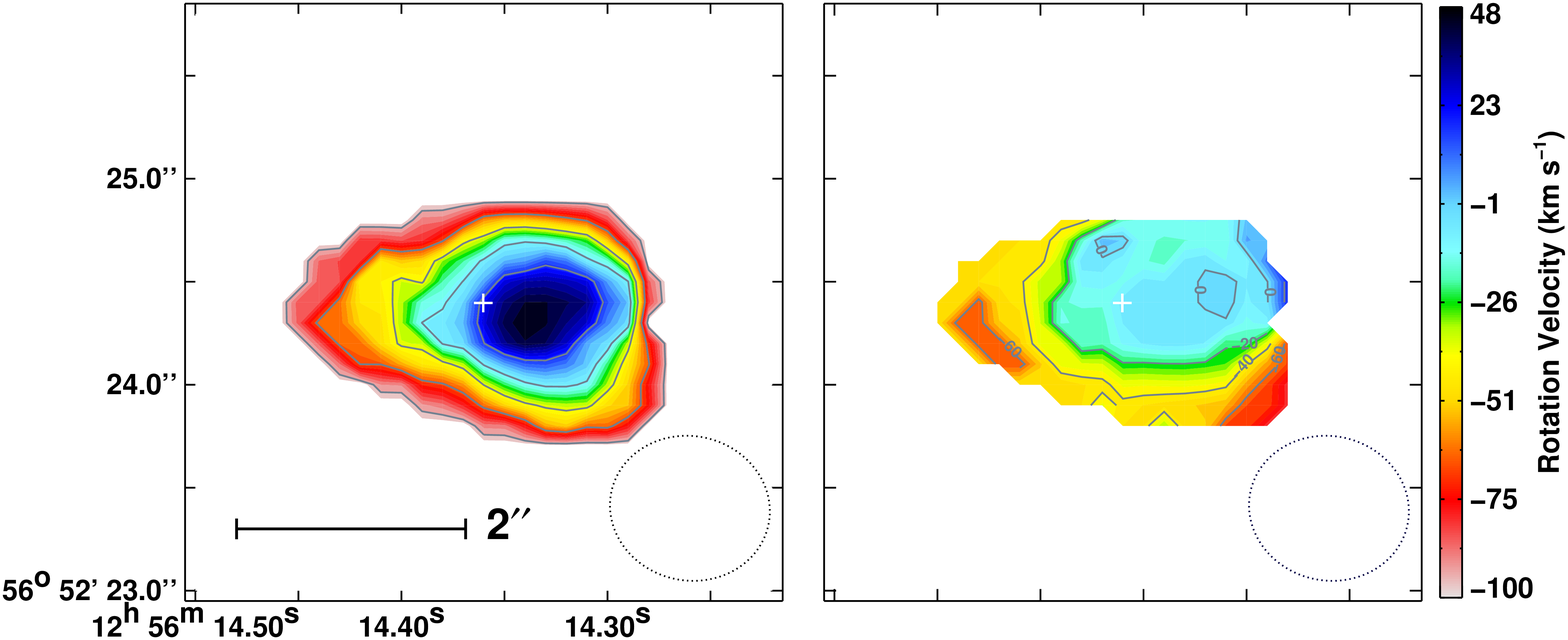}}
\subfigure{\includegraphics[height=4.8cm,clip,trim=1cm 0.5cm 0cm 3cm]{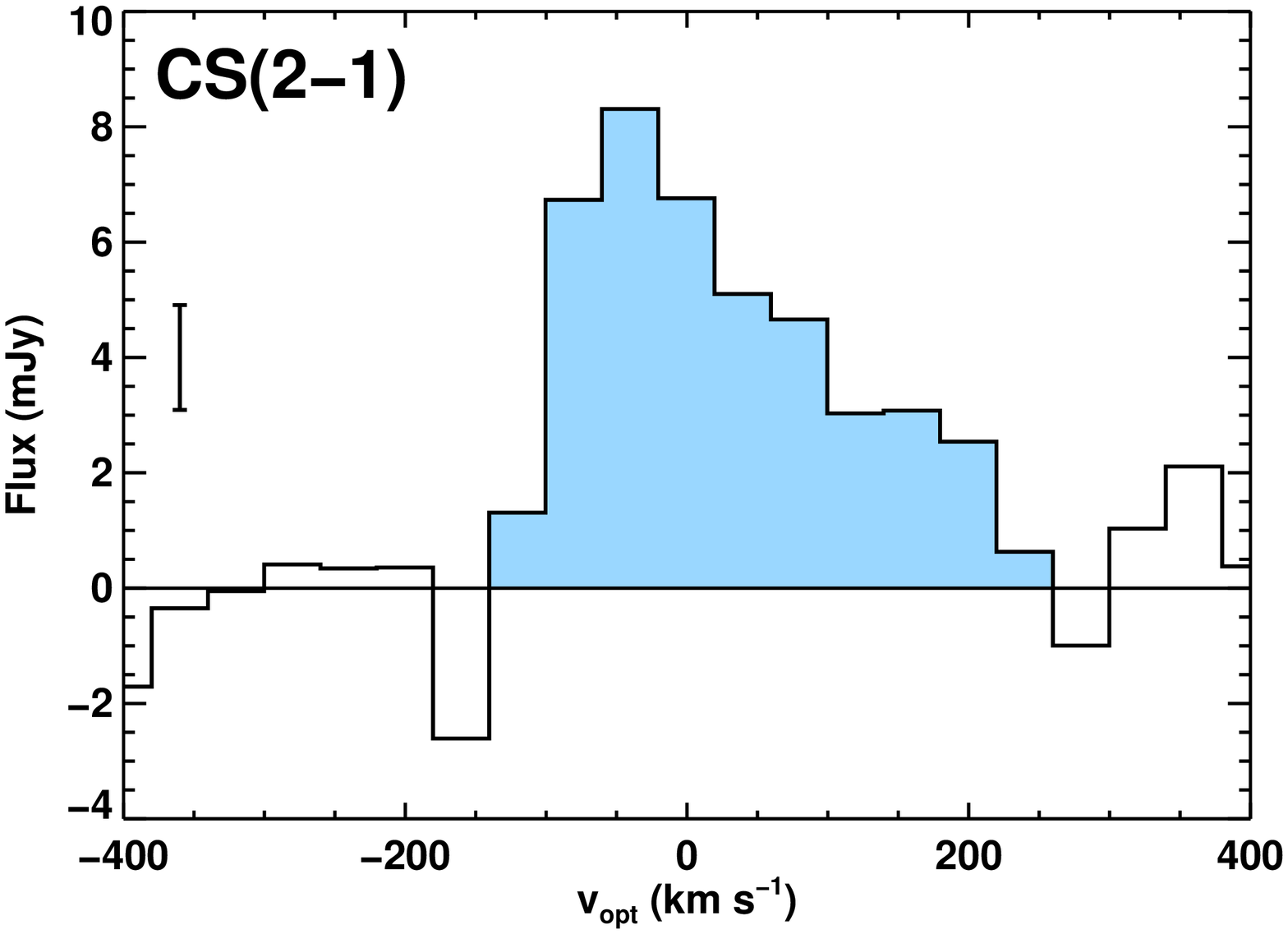}}
\subfigure{\includegraphics[width=0.99\textwidth]{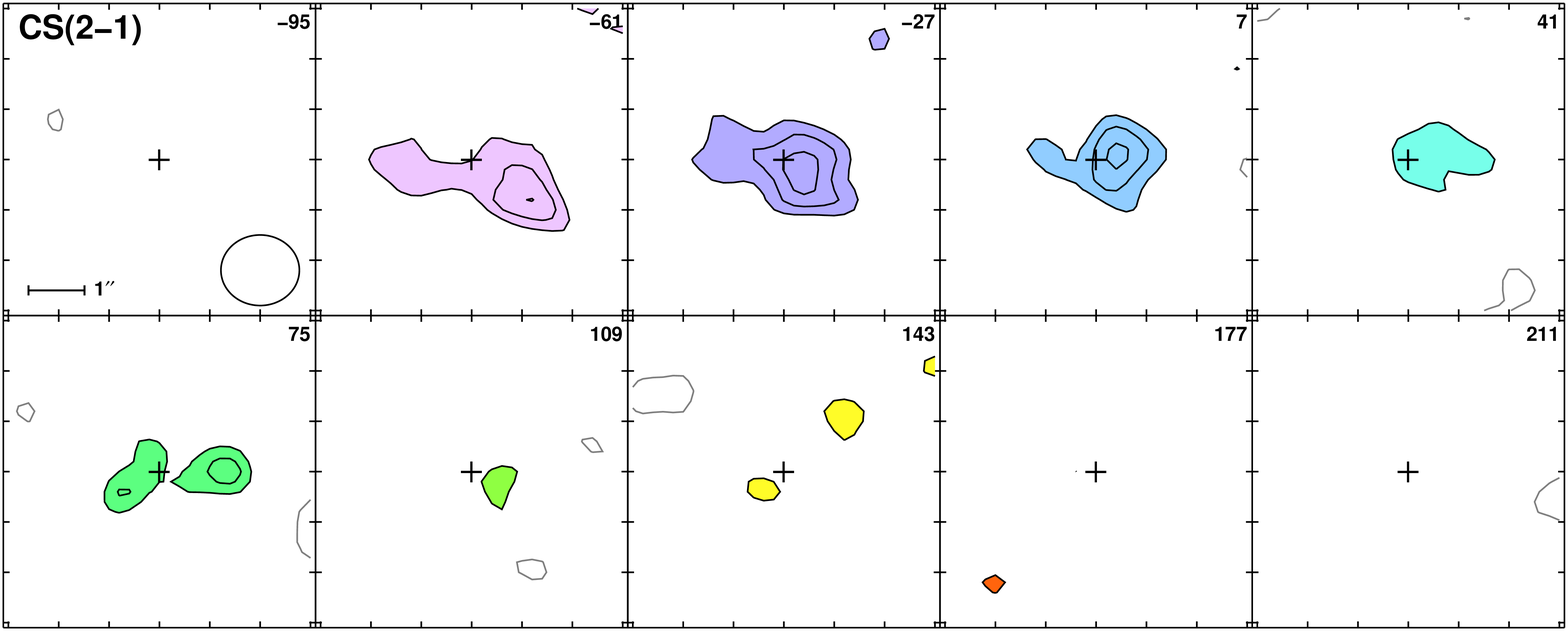}}
\caption{ {\bf(Top left \& middle):} The integrated intensity (moment0) and mean velocity (moment1) maps of CS(2--1) within Mrk\,231 from CARMA.  While there does appear to be some ordered motion, overall the spectrum appears quite turbulent.  Contour levels are [0.01,0.1,0.3,0.5,0.7,0.9] times the maximum value of the moment0 map. {\bf(Top right):} The integrated CS(2--1) spectrum, with the RMS per channel shown as a small error bar on the left-hand side of the spectrum.  The CS(2--1) line appears to have a width of $\approx$300\,\kms. {\bf(Bottom):} The CS(2--1) channel maps.  Contours are [-2.5,2.5,3.5,4.5]$\sigma$, where $\sigma$ is the rRMS of the CS(2--1) cube.  The negative contours are gray.}
\label{fig:cs}
\end{figure*}

CS(2--1) observations are shown in Figure \ref{fig:cs}.  Construction of channel maps, integrated intensity (moment0) maps, mean velocity (moment1) maps, spectra, and the corresponding determinations of the root mean square noise were performed identically to the methods described in \citet{alatalo+13}.  CS(2--1) channels have an (optically-defined) velocity width of 34\,\kms.  The total CS(2--1) flux, determined by summing across the shaded channels in the spectrum, is $1.43\pm0.19$\,Jy~km~s$^{-1}$, and covers an area of 3.9\,arcsec$^2$.

\section{Results and Discussion}
\label{disc}

\subsection{Dense molecular gas traced by CS(2--1)}
Using the local thermal equilibrium $F_{\rm CS(2-1)}$--$M_{\rm H_2,dense}$ relation discussed in \citet{a14_sfsupp} of 
\[N_{\rm CS} = 1.90\times10^{11} T_{\rm ex}~e^{7.05/T_{\rm ex}}~I_{\rm CS(2-1)}\]

\noindent where $I_{\rm CS(2-1)}$ is the integrated intensity of the CS(2--1) line in K~km~s$^{2-1}$, assuming $T_{\rm ex}\approx70$\,K \citep{vanderwerf+10}, and a CS abundance of $\approx10^{-9}$, we find a total dense gas mass of $\approx1.8\pm0.3\times10^{10}~M_\odot$.  This dense gas mass is consistent with the total dense gas mass found with other tracers \citep{solomon+92,solomon+vandenbout05,feruglio+10}, given the uncertainty in the CS/H$_2$ abundance, which is at least a factor of 2. Our dense gas mass estimate is lower than the dynamical mass derived within 1100\,pc of the center of Mrk\,231 is $3.15\times10^{10}$~M$_\odot$\footnote{Not including the uncertainty associated with converting CS to $M$(H$_2$) \citep{a14_sfsupp}} \citep{downes+solomon98}, necessitating that the majority of the central mass in Mrk\,231 is in the form of dense molecular gas. The signal-to-noise ratio of the CS data and resolution of these data are not able to confirm whether the CS is part of the face-on warped disk reported in \citet{davies+04}, and in dense gas by \citet{aalto+15}. Despite the limitation in resolution, the gradient seen in the CS moment1 map also appears consistent with the HCN kinematics seen by \citet{aalto+12}.   The agreement of the molecular gas mass also appears consistent with the idea that the majority of the molecular gas in Mrk\,231 is in a dense form.  

Our integrations were not sufficiently sensitive to detect CS(2--1) emission associated with broad wings (as is the case in other dense gas tracers, including HCN, reported by \citealt{aalto+12}), and warrants deeper observations to determine whether gas traced by CS (suggested to be a tracer of some of the densest molecular cores; \citealt{baan+08}) is also taking part in the molecular outflow.

To derive the dense gas column along the line-of-sight to the AGN in Mrk\,231 (white cross on Fig. \ref{fig:cs} moment maps), we find that the pixel associated with the radio point source contains intensity of $51.8\pm2.0$\,K~km~s$^{-1}$ (using a Kelvin per Jansky factor of 78.35).  Converting this intensity into a column density of H$_2$ (assuming an excitation temperature of 70\,K and a CS/H$_2$ abundance of 10$^{-9}$), we find a CS-derived line-of-sight column to the AGN of $1/2\times N_H=N({\rm H_2})\approx1.1\times10^{24}$\,cm$^{-2}$ (if the AGN were sitting beneath the maximum value in the CS(2--1) map, the obscuring column would be $N({\rm H_2})\approx1.5\times10^{24}$\,cm$^{-2}$). While this value was consistent with a previous estimate from X-ray absorption models \citep{braito+04}, newer results from {\em NuSTAR} show $N_H$ to be $1.2\times10^{23}$~cm$^{-2}$ \citep{teng+14}.  The discrepancy between the CS-derived column might suggest that the obscuring medium in the nucleus of Mrk\,231 is quite clumpy (at a scale much smaller than the CARMA resolution), which is consistent with the suggestion by \citet{teng+14} that the wind punching out of the system has created a preferential line-of-sight to the AGN.

If we were to use the CS to investigate the star formation efficiency within Mrk\,231, the total dense gas-traced surface density is $\Sigma_{\rm H_2} \approx 6450~M_\odot$~pc$^{-2}$.  Comparing this to the star formation surface density (assuming the dense gas and star formation are co-spatial) of $61~M_\odot$~yr$^{-1}$~kpc$^{-2}$, we find that this matches the expectation from the Kennicutt-Schmidt relation \citep{ken98}.

\subsection{Resolved CO(1--0) outflow}
Accounting for the distance to Mrk\,231, $0.49''$ corresponds to a projected separation between the CO(1--0) centroids of 415\,pc. The CARMA data also successfully resolve the lobes, with sizes of 2.2$''$ and 2.3$''$ of the red- and blue-shifted wings, respectively.  To deproject this distance, an accurate inclination angle is needed.  Inclination angles derived for the molecular disk (which we assume is perpendicular to the outflow) has derived inclinations angles ranging from $i=10^\circ$ \citep{downes+solomon98,davies+04} to $i=45^\circ$ \citep{richards+05}, meaning that the deprojected distance ranges between 570 and 2390\,pc.  If we assume that the outflow is conical with a large opening angle \citep{rupke+11,cicone+12}, then the   distance between the wings is likely similar to the diameter of the lobes, which we estimate to be 1800\,pc (after de-convolving the lobes).   This size is slightly larger than the CO(1--0) separation found by \citet{cicone+12}, although we see extension in the East-West direction and opposite of the rotation of the CS disk, rather than the North-South direction as was observed in HCN by \citet{aalto+12}.  While this is consistent with the angle of rotation of the molecular disk, the velocities probed by the CARMA observations exceed the escape velocity, and therefore are unlikely to be due to rotation.

We use 600\,\kms\ for the characteristic velocity of each wing (as well as the de-projected radius, 900\,pc) to calculate the dynamical timescale of the outflow of $\tau_{\rm dyn} \approx 1.5$\,Myr, about twice as long as the timescale originally reported by \citet{feruglio+10}.  If we use this dynamical timescale in conjunction with the outflow mass from \citet{feruglio+10}, and supported by \citet{aalto+12} of 5.8$\times10^8~M_\odot$, we find that the total mass outflow rate is $390~M_\odot$~yr$^{-1}$, a factor of 2 smaller than \citet{feruglio+10}, and a factor of $\approx2$ larger than the star formation rate in the system \citep{veilleux+09}.  If all $390~M_\odot$~yr$^{-1}$ taking part in the molecular outflow were to {\em escape} the galaxy, we would expect the molecular gas to be depleted in $\approx46$\,Myr through the action of the molecular outflow alone, and 32\,Myr if both outflow and the star formation gas consumptions are considered.  This depletion timescale assumes that all molecular gas in the thin disk is intercepted by the molecular outflow, which is assumed to be traveling perpendicular to the disk.  If the molecular outflow is unable to interact with most of the gas in the molecular disk, this could increase the molecular outflow depletion timescale considerably.

\subsection{The importance of competing depletion mechanisms}
In Mrk\,231, the molecular gas is being depleted by three competing mechanisms: consumption through star formation, escape from the galactic potential via a molecular outflow and accretion onto the supermassive black hole.  The timescale for star formation to completely consume the central gas is 110\,Myr.  The molecular outflow appears to be the dominant depleter within the Mrk\,231 system (assuming that the outflowing molecular mass is being completely expelled). 

While $\dot{M}_{\rm out}$ of Mrk\,231, updated in the prior section is indeed larger than $\dot{M}_{\rm SFR}$, $\dot{M}_{\rm out}$ might not be an accurate representation of the mass of molecular gas that is currently being expelled from the system.  Mrk\,231 is a massive system, with an estimated stellar mass of $\approx3\times10^{11}~M_\odot$ \citep{U+12}, and thus much of the outflowing molecular gas will not escape the potential well of Mrk\,231, instead falling back and ``stirring up'' the massive molecular disk in the center.  
Using a dynamical mass within 1100\,pc of the center of Mrk\,231 of $3.15\times10^{10}~M_\odot$ \citep{downes+solomon98,davies+04}, and derive an escape velocity from the central kiloparsec of $\sim500$\,\kms.  If we assume that all gas that is traveling above $v_{\rm esc}$ will be able to successfully escape the system, then 15\% of the total outflowing molecular gas (from the gaussian fit of \citealt{feruglio+10}) escapes\footnote{This estimate only accounts for the the molecular gas traced by CO(1--0), although the Mrk\,231 outflow has been detected in additional tracers of netural gas, including OH \citep{sturm+11}, Na\,D \citep{veilleux+09} and \hi\ \citep{teng+13}.}.  We use our updated maps to calculate the total escaping mass from the blue- and red-shifted wing fluxes reported in Section \ref{obs} and Table \ref{tab:results}, using a conservative merger-based conversion factor \citep{narayanan+11}, and find $M_{\rm esc} = (2.93\pm0.49)\times10^8~M_\odot$.  Dividing by the dynamical time, we find
 $\dot{M}_{\rm esc}\approx 195~M_\odot$~yr$^{-1}$, which is quite comparable to the current star formation rate of 170~$M_\odot$~yr$^{-1}$ than were we to assume all outflowing mass is escaping.  This updates the depletion timescale due to the escape of molecular gas, $\tau_{\rm esc}$ to 92\,Myr.  The combined depletion timescale $\tau_{\rm SF+esc}$ is then 49\,Myr.

The accretion onto the black hole can be derived using the bolometric luminosity of the AGN (of $\approx 2.8\times10^{12}~L_\odot$; \citealt{veilleux+09,veilleux+13}), and assuming an efficiency $\eta = 0.17$, to be $\dot{M}_{\rm acc}\approx 1~M_\odot$~yr$^{-1}$, or at a rate $\sim10^{-2}$ other depletion rates, a minor constituent to the depletion of the central molecular gas, and requiring a Hubble time to completely remove the molecular gas from the system.

Without a constant driving mechanism, the majority of the molecular outflow will eventually fall back into the center, either to be accreted, formed into stars, or re-launched.  One would expect that this gas would fall back and re-inject energy into the central molecular disk, possibly acting to inhibit star formation \citep{guillard+14,a14_sfsupp}, and therefore extend the molecular gas consumption timescale.  An extension of the depletion timescale expected for Mrk\,231 from $\sim20$\,Myr (the typical lifetime for O-stars) to $\sim50$\,Myr, appears more consistent with the poststarburst population found in Mrk\,231 \citep{canalizo+00}, and common in other quasars \citep{cales+11,canalizo+13}, as well as other hosts of molecular outflows (such as NGC\,1266; \citealt{alatalo+11,a14_stelpop}).

The rapidity at which Mrk\,231 is currently expelling its molecular material compared with weaker AGNs might be due to the different circumstances that created this system and outflow. A major merger occurred in a system already sufficiently massive to host a massive black hole, which is thus energetically capable of driving the bulk of the molecular gas from the system.  Sub-$v_{\rm esc}$ gas would be able to fall back into the system and inject turbulence, in effect, temporarily stalling efficient star formation and possibly extending the depletion time just slightly.  This would account for the mismatch in depletion timescales between $\sim30$\,Myr (the depletion timescale should all outflowing gas escape and with the current star formation rate) and $\sim50$\,Myr (the stellar population age seen in Mrk\,231; \citealt{canalizo+00}).

Deeper observations of the molecular outflow in Mrk\,231, focusing on spatial resolution on at different distances away from the AGN will be able to determine what fraction of the molecular gas is escaping the gravitational potential, and provide a more complete understanding of the driving mechanism as well as the energy injection rate.  Low frequency imaging could pinpoint the location of fossil shells \citep{schoenmakers+00}, as Mrk\,231 has already been detected at low frequency \citep{cohen+07}.  Combining these observations with the multiwavelength suite available for Mrk\,231 will provide additional constraints, such as whether the molecular outflow changes phase as it accelerated, and whether there is any evidence that gas ultimately falls back into the molecular disk of Mrk\,231.

\section{Conclusions}
\label{conc}
We have presented 3mm continuum, and CO(1--0) and CS(2--1) high resolution molecular gas maps of Mrk\,231 from CARMA.  The CS(2--1) data paint a picture of the dense molecular gas in Mrk\,231 that is consistent with other dense gas tracers \citep{aalto+12}.  CS(2--1) predicts a dense gas mass (of $M_{\rm dense}=1.8\pm0.3\times10^{10}~M_\odot$), and a line-of-sight column toward the AGN of $N({\rm H_2}) \approx 10^{24}$~cm$^{-2}$, inconsistent with the X-ray derived column \citep{teng+14}, possibly indicative of a clumpy ISM, and the AGN-driven winds punching a preferential line-of-sight to the AGN.  If we use the dense gas to derive a star formation efficiency, we find that Mrk\,231 is consistent with the prediction of the Kennicutt-Schmidt relation.  

CARMA was able to resolve the individual CO(1--0) broad line-wings, with an inferred separation between the wings to be $\approx1.8$\,kpc given the angular size of the lobes and possible inclination angles of the outflow.  This is slightly larger than the CO(1--0) lobe separation reported in \citet{cicone+12}.  If we assume that not all molecular gas will successfully escape the galaxy, we can re-calculate the depletion timescale for Mrk\,231 based on the mass {\em escape} rate, of $\approx195~M_\odot$~yr$^{-1}$.  A depletion timescale using $\dot{M}_{\rm esc}+\dot{M}_{\rm SFR}$, rather than $\dot{M}_{\rm out}+\dot{M}_{\rm SFR}$ is more consistent with the poststarburst stellar populations that are seen in Mrk\,231 as well as many other quasars.

\section{Acknowledgments}
K.A. would like to thank Mark Lacy, Lauranne Lanz and Kristina Nyland for useful conversations and proofreading, which improved the text, as well as the anonymous referee, for a substantive report that has significantly improved the manuscript.
K.A. is supported by funding through Herschel, a European Space Agency Cornerstone Mission with significant participation by NASA, through an award issued by JPL/Caltech.  Support for CARMA construction was derived from the states of California, Illinois, and Maryland, the Gordon and Betty Moore Foundation, the Kenneth T. and Eileen L. Norris Foundation, the Associates of the California Institute of Technology, and the National Science Foundation. Ongoing CARMA development and operations are supported by the National Science Foundation under a cooperative agreement, and by the CARMA partner universities.  This research has made use of the NASA/IPAC Extragalactic Database (NED) which is operated by the Jet Propulsion Laboratory, California Institute of Technology, under contract with the National Aeronautics and Space Administration.  Based on observations made with the NASA/ESA Hubble Space Telescope, and obtained from the Hubble Legacy Archive, which is a collaboration between the Space Telescope Science Institute (STScI/NASA), the Space Telescope European Coordinating Facility (ST-ECF/ESA) and the Canadian Astronomy Data Centre (CADC/NRC/CSA).
\\
\bibliographystyle{apj}
\bibliography{ms}

 \end{document}